\documentclass[12pt]{article}

\usepackage{epsf}

\begin{document}

\begin{center}
{\bfseries \Large Heisenberg model of the high-energy hadron
collision in terms of chiral fields}

\vskip 5mm

Oleg V. Pavlovsky$^{\dag}$

\vskip 5mm

{\small {\it Intitute for Theoretical Problems of Microphysics,
Moscow State University, Moscow, Russia }
\\
$\dag$ {\it E-mail: ovp@goa.bog.msu.ru }}
\end{center}

\vskip 5mm

\begin{center}
\begin{minipage}{150mm}
\centerline{\bf Abstract} Properties of chiral Born-Infeld Theory
proposed as the model for shock-wave fireball production in the
hadron-hadron collisions was studied. The role of the shock-waves
in the multi-particle production was discussed.
\end{minipage}
\end{center}

\vskip 10mm

\section{Introduction and general motivation}

In spite the quantum chromodynamics being considerably successful
for last 30 years, many opened questions are still waiting for
their solution. It is now obvious that the full quantitative
theory of strong interactions should be derived from the
fundamental QCD Lagrangian and must take into account the most
important features of the QCD vacuum, such as the confinement and
the chiral symmetry breaking. A number of lattice data, e.g., the
disappearance of the chiral condensate and the confinement at the
same temperature, is known to point out the deep internal
causation between these phenomena.

Another great open problem of the modern particle physics is a
saturation of the Froissart bound at high-energy hadron collision.
And, as we will see, this problem surprisedly connects very
closely to the first one.

Last years (due to the investigation of quark-gluon plasma) a
great interest arose to the problem of multi-particle production
and the Froissart bound saturation of total cross-sections, first
considered by Heisenberg 50 years ago \cite{heisenberg}.

There are in the order of 3000 charged particles emerging from
central collisions at high-energy experiments in RHIC and most of
them are pions. It is a very attractive idea to describe this
multiple-pion emission in high-energy hadronic and/or nuclear
collisions as classical radiation of $\pi$-meson field
\cite{heisenberg}.  This idea has been rediscovered and further
developed in the modern context of low-energy effective theories
\cite{anselm, blaizot}. On the other hand, the fireball production
problem in Heisenberg picture has been considered in resent papers
\cite{nastase} in the framework of AdS-CFT correspondents. All
these results in collection with new experimental data have been
stimulating investigation on this subject.

In this work we am going to reformulate Heisenberg's ideas in
terms of chiral fields and to find a connection between the
concept of the shock-wave fireball production and QCD.
 The ``shock-wave''-like solutions (analogues of the 1+1D solutions in
the Heisenberg model) of the Chiral Born-Infeld theory will be
studied.

\section{Chiral Bag solution of Chiral Born-Infeld Theory}

For studding of the process of high-energy hadron collision the
idea of fireball production is commonly used. The fireball is a
state of the nuclear matter with very high density. Such state
usually is simulated by Chiral Bag with quasi-independent quarks
and gluons inside. Heisenberg \cite{heisenberg} was a first who
proposed the model where such object as fireball naturally arises
due to shock-wave phenomena.

Heisenberg model is based on Born-Infeld Lagrangian for scalar
fields $\phi$:
\begin{equation}
{\cal L}_{MesonBI} = - \beta^2 \bigg(1-\sqrt{1-\frac
1{2\beta^2}\phi_{,\mu} \phi^{,\mu} } \bigg)  . \label{2.0}
\end{equation}
One should note that this lagrangian can be reformulated as
lagrangian of 3D Bosonic String Theory \cite{barbashov} and now
this model very widely used in cosmology \cite{BIcosmology}.

The starting point of our consideration is a idea that the
fireball at the first moment of the central collision (the state
with high density of the nuclear matter) can be treated as the
chiral soliton analogously to the nucleon soliton model. Such a
state must be very unstable and quickly decay into secondary
particles and the Heisenberg model describe this process very
well.

Let us consider a direct analogue of the Heisenberg Lagrangian for
chiral field
\begin{equation}
{\cal L}_{ChBI} = - f^2_\pi {\rm Tr} \beta^2 \bigg(1-\sqrt{1-\frac
1{2\beta^2}L_\mu L^\mu } \bigg) \stackrel{\beta \to \infty
}{\longrightarrow} - \frac{f^2_\pi}{4} {\rm Tr} L_\mu L^\mu ,
\label{2.1}
\end{equation}
where $\beta$ is the mass dimensional scale parameter of our
model. It can be easily shown that the expansion of the Lagrangian
(\ref{2.1}) gives us the prototype Weinberg theory as the leading
order theory in the parameter $\beta$ and Heisenberg theory as the
leading order theory in the parameter $f_\pi$ . Moreover this is
topologically non-trivial theory and one can identify the
topological number of solution with baryon number as usual. Now we
consider the spherically symmetrical field configuration
\begin{equation}
U=e^{i F(r) (\vec n \vec \tau)}, \qquad \vec n = \vec r/|r| .
\label{2.2}
\end{equation}

Using the variation principle, we get the equation of motion
$$
(r^2 - \frac 1{\beta^2} \sin^2 F) F'' + ( 2rF' -\sin 2F) -
$$
\begin{equation}
- \frac 1{\beta^2} (r F'^3 - F'^2 \sin 2F + 3 \frac 1{r} F'
\sin^2F - \frac 1{r^2} \sin2F \sin^2 F)=0. \label{2.4}
\end{equation}

Equation (\ref{2.4}) has a singular region (singular surface) with
singular behavior of solutions. Let $ r_0 $ belong to such
singular surface. Then
\begin{equation}
\left \{  \begin{array}{rcl}
            (\beta r_0)^2  - \sin^2 F_0 &=& 0\\
F'_0 \Bigl( (F'_0)^2 \sin F_0 \mp F'_0 \sin 2F_0 + \sin^2 F_0
 \Bigr) &=& 0 \, \, \, \, \mbox{where} \, F_0=F(r_0) .
           \end{array}
\right. \label{2.61}
\end{equation}
Equations (\ref{2.61}) have only two solutions: whether
\begin{equation}
\left \{  \begin{array}{lcl}
           r_0 &\neq& 0 , \\
           F_0 &=& \pm \arcsin(\beta r_0) + \pi N   , \, \, \, \mbox{where} \,
 N \in \mbox{\textbf{Z}}, \\
          F'_0 &=& 0.
          \end{array}
\right. \label{2.62}
\end{equation}
or
\begin{equation}
\left \{  \begin{array}{lcl}
           r_0 &=& 0 , \\
           F_0 &=&  \pi N   , \, \, \, \mbox{where} \, N \in
           \mbox{\textbf{Z}}, \\
           F'_0 &\neq& 0.
           \end{array}
\right. \label{2.63}
\end{equation}

Topological solitons of ChBI theory  correspond to the possibility
(\ref{2.63}. In my talk I consider the more interesting
possibility (\ref{2.62}).

One obtains the asymptotic behavior near the singular surface ($r
\to r_0$, $F(r \to r_0) \to \arcsin(\beta r_0)$)
\begin{equation}
F(r \to r_0) = \arcsin(\beta r_0) + \mbox{\textbf{b}} (r -
r_0)^{3/2} + \underline{O}((r - r_0)^{2}), \label{2.64}
\end{equation}
where $\mbox{\textbf{b}}$ is a constant. Of course, the derivative
$F'_0$ at the point $r=r_0$ is zero.

 \begin{figure}[t]
 \leavevmode
 \epsfbox{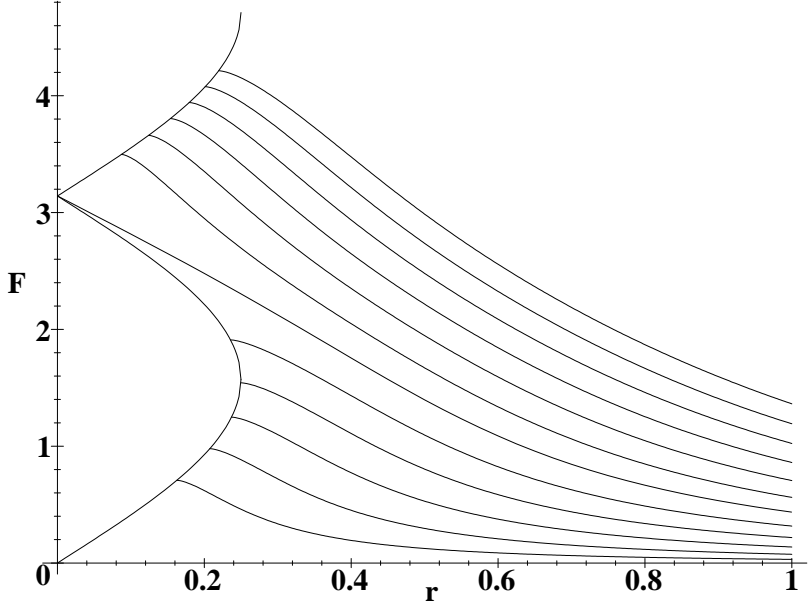}
 \caption{ Solutions of equation (9)
which have the asymptotics (15) ($a_\infty>0$) at infinity.
Horizontal axis: r (in fm).}
 \end{figure}

The numerical investigation of the solutions of equation
(\ref{2.4}) is presented in Fig.1. Most of these solutions can be
evaluated only for $r>r_0$, where $r_0$ is determined by $F(r_0)=
\pm \arcsin (\beta r_0)$. But among this set of solutions there
are solutions with $r_0=0$. Such solutions have the asymptotics
\begin{equation}\label{2.66}
F(r) = \pi N  + a r - \frac{7a^2-4\beta^2}{30(a^2-\beta^2)} a^3
r^3+ \underline{O}(r^{5})
\end{equation}
at origin ($a^2 < \beta^2/3$ is a constant), and these are the
topological solitons of the ChBI theory. The scale parameter
$\beta=807 \, \mbox{MeV}$ is preliminarily defined from the
hypothesis that the soliton with $B=1$ is a nucleon.

Now we would like to draw the attention to another class of
solutions. These solutions are defined everywhere, except the
small ($ \sim$ 0.2 fm) spherical region about the origin. These
solutions look like a ''bubble" of vacuum in the chiral fields and
are of the interest for the chiral bag model. In the internal
region ($r<r_0$), the only vacuum configuration can exist. From
the mathematical point of view, such "step-like'' solutions with
jump of $F(r)$ at $r=r_0$ are a generalized solutions.

 \begin{figure}
 \leavevmode
 \epsfbox{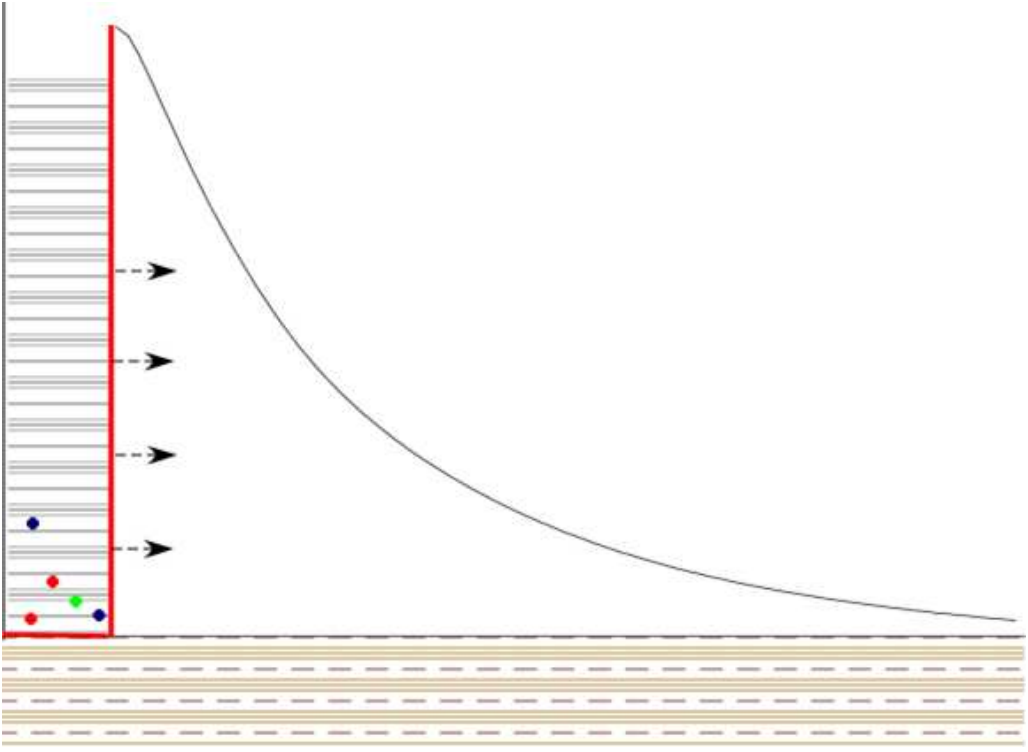}
 \caption{Decay of ChBI fireball.}
 \end{figure}

To clarify the physical nature of this "step-like'' generalized
solutions, let us consider the projection of the left-hand chiral
current on the outward normal of the singular surface. The
singular surface of "step-like'' solutions is a confinement
boundary surface for constituent quarks. For the self-consistency
of our "two-phase" picture, let us check is it possible to
compensate the non-zero quark chiral current on the confinement
surface by the non-zero chiral current of ChBI "step-like''
configuration that appear due to the defect on the singular
surface. The projection the chiral current on the outward normal
of the singular surface for the spherically symmetrical
configuration (\ref{2.2}) reads
\begin{equation} \label{2.71}
(\vec n \, \vec J_\pi )|_{\partial V}= f^2_\pi {\rm Tr}
\frac{\tau^a  \vec x  \vec L }{\sqrt{1-\frac 1{2\beta^2}L_\mu
L^\mu }} \, = \, \frac{3}{2} f^2_\pi \frac{\mbox{\textbf{b}}
r_0}{\sqrt{2 r_0 + 9 r_0^2 \mbox{\textbf{b}}^2 }},
\end{equation}
where the coefficient $\mbox{\textbf{b}}$ from the asymptotic
(\ref{2.64}) is a function of $F_0=F(r_0)$ and can be evaluated
numerically.

The crucial point for such analysis  steams  from the fact that
$\mbox{\textbf{b}}(F_0)$ has the singularity at $F_0=\pi N$ and
$r_0=0$, or $\mbox{\textbf{b}}(F(r_0=0))=\infty$. This implies
that the soliton solutions of the ChBI theory are solutions with
point-like singular chiral source and the "bag"-like solutions are
the solutions with the some distribution of the chiral current on
the confinement surface. It is possible to show that for any
internal constituent quark configuration with confinement inside
some volume $V$ the solution of ChBI theory $U( \vec \pi)$ could
be defined which compensate the non-zero quark's chiral current
across the surface $\partial V$
\begin{equation}\label{2.72} (\vec n \, \vec J_q )|_{\partial V} =
\sum_g{\frac{i}{2}\bar{\Psi}_q (\vec{x}\vec{\gamma}) \gamma_5
\Psi_q} = (\vec n \, \vec J_\pi )|_{\partial V}.
\end{equation}
Equation (\ref{2.72}) can be considered as a condition on
coefficient $\mbox{\textbf{b}}(\partial V)$, and plays the role of
the self-consistency condition between quark and chiral phases.

The qualitative picture of the decay of high density Chiral
Born-Infeld soliton (ChiBIon) is follows. As it was shown in
\cite{Pavlovsky:2004np} the ChiBIon with large topological charge
is very unstable ($E_{{ChiBIon}} \sim B^3$ for the topological
charge $B \gg 1$). It is obvious that the most energetically
profitable decay is going on via the step-like singularities
production (see  in Fig. 2). Such singularities should excite the
chiral degrees of freedom from vacuum inside the "bubble" and the
ChiBIon should explode very quickly.

\section{Conclusions}
The aim of this paper is to study of  ``step-like'' generalized
solutions of the Born-Infeld theory for chiral fields. Physically
the critical behavior in the ChBI theory appears when the chiral
field strength of the prototype field theory approaches the value
of the squared effective coupling constant ($\beta^2$).

The Chiral Born-Infeld theory is a good candidate on the role both
for the model for the chiral cloud of the baryons and for the
fireball. One can show that energy of Chiral Born-Infeld
topological solitons grows up $B^3$ for large topological numbers
$B$. It's means that any compact clusters with a big baryon number
are very instable and quickly decay on individual baryons and
mesons. The decay of such state is a very stochastic process and
leads to generation of large number of hadrons. Details of this
process would be the topics for future works.

This work is partially supported by the Russian Federation
President's Grant 4476-2006-2. The hospitality and financial
support of the ECT* in Trento is gratefully acknowledged.

\end{document}